\documentstyle[aps,prd,preprint]{revtex}
\tighten
\begin{document}
\draft

\preprint{\vbox{\baselineskip=12pt
\rightline{gr-qc/9412073}
\rightline{Submitted to Physical Review D}}}

\title{The Evolution of Distorted Rotating Black Holes II:
Dynamics and Analysis}

\author{Steven R. Brandt
 and Edward Seidel}

\address{ National Center for Supercomputing Applications \\
605 E. Springfield Ave., Champaign, Illinois 61820}

\address{ Department of Physics,
University of Illinois at Urbana-Champaign, Urbana, Illinois 61801}

\date{\today}
\maketitle

\begin{abstract}
We have developed a numerical code to study the evolution of
distorted, rotating black holes.  This code is used to evolve a new
family of black hole initial data sets corresponding to distorted
``Kerr'' holes with a wide range of rotation parameters, and distorted
Schwarzschild black holes with odd-parity radiation.  Rotating black
holes with rotation parameters as high as $a/m=0.87$ are evolved and
analyzed in this paper. The evolutions are generally carried out to
about $t=100M$, where $M$ is the ADM mass.  We have extracted both
the even- and odd-parity gravitational waveforms, and find the
quasinormal modes of the holes to be excited in all cases.  We also
track the apparent horizons of the black holes, and find them to be a
useful tool for interpreting the numerical results.
We are able to compute the masses of the black holes from the
measurements of their apparent horizons, as well as the total energy
radiated and find their sum to be in excellent agreement with the ADM
mass.
\end{abstract}


\narrowtext

\section{Introduction}
\label{intro}

In this paper we present the first results for the evolution of
initial data sets corresponding to axisymmetric, rotating vacuum black
holes. We developed a numerical scheme and code based on an extension
of earlier work at
NCSA~\cite{Bernstein93b,Bernstein94a,Abrahams92a,Anninos93c} on
distorted, but non-rotating, axisymmetric black holes.  In a companion
paper, referred to henceforth as Paper~I~\cite{Brandt94b}, we describe
the numerical methods, gauge conditions and code tests developed for a
code to evolve rotating black hole spacetimes.  With this new code we
have been able to evolve and study a new family of distorted rotating
black hole data sets.  The construction of these data sets was
outlined in Paper~I, and will be discussed in another paper in this
series~\cite{Brandt94a}, which we refer to as Paper~III.  In this
paper we focus on analyzing the physics of the evolution for a number
of different rotating black hole data sets, with both even- and
odd-parity radiation.  We also consider the evolution of non-rotating
black holes that have been distorted by the presence of odd-parity
gravitational waves.

These rotating black holes can be highly distorted, allowing one to
study their nonlinear dynamics.  These studies will be useful not only
in understanding rotating black hole spacetimes, but also in studying
the late stages of black hole collisions with angular momentum, just
after the holes have coalesced.  At that time they will have formed a
single, highly distorted rotating hole similar to the configurations
studied here.

In analyzing these rotating black hole evolutions, we have developed a
series of tools that allow us to study gravitational waves, apparent
horizons, and other quantities of interest.  Using these tools we have
extracted the waveforms for both the even- and odd-parity radiation
emitted by a distorted rotating black hole, and we find that the
quasinormal modes of the hole are excited.  The extracted waveforms
are also used to compute the energy carried away from the black hole
via gravitational waves.  We locate and study the geometry of the
apparent horizons in these spacetimes, and find that their geometric
structure can be used as a key tool in analyzing the evolution.  We
show how to extract information about the oscillation frequency, the
mass, and the angular momentum of the black hole.

The organization of this paper is as follows: In
section~\ref{preliminaries} we define the variables used in our
simulations and describe the variety of initial data sets we evolve
and analyze in this paper.  In section~\ref{nzp} we present an
analysis of ``near zone'' results from evolving various initial data
sets, including an analysis of the physics that can be extracted from
apparent horizons.  Section~\ref{farzone} contains a discussion of our
extraction methods to compute the even and odd-parity radiation in the
spacetime, while section~\ref{Results} contains a discussion of the
results of evolving the data sets.  Finally, in section~\ref{conc} we
conclude and discuss future directions this research will take.

\section{The Code and Initial Data}
\label{preliminaries}

The code we developed to evolve rotating black hole data sets has been
discussed in detail in Ref.~\cite{Brandt94b}.  We will not discuss any
details of the numerical methods used to evolve the spacetimes here,
but for context we define our notation for the primary variables used
in our code.

\subsection{Definition of Variables}
\label{dov}
We build on earlier work of Ref.~\cite{Bernstein93b,Anninos93c} on
non-rotating black holes in defining the variables used in our code.
Additional metric and extrinsic curvature variables must be introduced
to allow for the odd-parity modes present in this new system.
Previously only even-parity modes were present. We define the
variables used in our evolutions as follows:
\begin{mathletters}
\begin{eqnarray}
\gamma_{ij} & = &\left( \begin{array}{ccc}
\gamma_{\eta\eta} & \gamma_{\eta\theta} & \gamma_{\eta\phi}\\
\gamma_{\eta\theta} & \gamma_{\theta\theta} & \gamma_{\theta\phi}
\\
\gamma_{\eta\phi} & \gamma_{\theta\phi} & \gamma_{\phi\phi}
\end{array} \right) =  \Psi^4 \left( \begin{array}{ccc}
A & C & E \sin^2\theta \\
C & B & F \sin\theta \\
E \sin^2\theta & F \sin\theta & D \sin^2\theta
\end{array} \right)
\end{eqnarray}
and
\begin{eqnarray}
K_{ij} & = & \Psi^4 h_{ij} = \Psi^4 \left( \begin{array}{ccc}
H_A & H_C & H_E \sin^2\theta \\
H_C & H_B & H_F \sin\theta \\
H_E \sin^2\theta & H_F \sin\theta & H_D \sin^2\theta
\end{array} \right).
\end{eqnarray}
\end{mathletters} \\
In these expressions $\eta$ is a logarithmic radial coordinate, and
($\theta$,$\phi$) are the usual angular coordinates.  The relation
between $\eta$ and the standard radial coordinates used for
Schwarzschild and Kerr black holes is discussed in Paper~I.  As in
Ref.~\cite{Anninos93c}, the conformal factor $\Psi$ is determined on
the initial slice and held fixed in time afterwards.  The introduction
of $\Psi$ into the extrinsic curvature variables simplifies the
evolution equations somewhat.  The various factors of $\sin\theta$ are
included in the definitions to explicitly account for their behavior
near the axis of symmetry and the equator, as discussed in Paper~I.

\subsection{Initial Data}
We constructed a new class of initial data sets corresponding to
distorted rotating black holes and odd-parity distorted non-rotating
black holes.  The construction of these data sets is described in
Paper~I and in more detail in Paper~III.  Here we briefly describe the
different classes of data sets and describe our parameter choices for
evaluation and analysis. There are three basic families of black hole
data sets that are evolved and analyzed in this paper.  The first
series of data sets, labeled in Table~\ref{tab:the_k's} as runs {\it k1}
and {\it k2}, correspond to Kerr black holes with rotation parameters
$a/m=.48$ and $.68$ respectively.  As described in Paper~I, these data
sets have been transformed from Boyer-Lindquist coordinates into our
logarithmic ``$\eta$'' coordinates and then evolved.

The second family of initial data sets that we consider are related to
the rotating black hole initial data of Bowen and York~\cite{Bowen80},
but we have added a ``Brill wave'' to distort them with a construction
similar to that of Ref.~\cite{Bernstein93a}.  As described in Paper~I,
the initial three-metric $\gamma_{ab}$ is chosen to be
\begin{equation}
ds^2=\Psi^4 \left[ e^{2 q} \left(d \eta^2+d\theta^2 \right)
+\sin^2\theta\, d\phi^2 \right].\label{mymet}
\end{equation}
where the ``Brill wave'' function $q$ is given by
\begin{mathletters}
\begin{eqnarray}
q &=& \sin^n\theta\, q_{G}, \\
q_{G} & = & Q_0 \left( e^{-s_+}+e^{-s_-} \right), \\
s_\pm & = & \left(\eta \pm \eta_0 \right)^2/\sigma^2.\label{metric form}
\end{eqnarray}\label{brill}
\end{mathletters} \\
The extrinsic curvature is computed following Bowen and York, with a
free parameter $J$ that specifies the angular momentum of the hole.
For more details of how the constraints are solved, please refer to
Paper~I.  Here we simply point out that the Bowen and York solution to
the momentum constraint is:
\begin{mathletters}
\begin{eqnarray}
H_E &=& 3 \Psi^6 J, \\
H_F &=& 0.
\end{eqnarray}
\end{mathletters}
We note that for $J=0$, the data sets reduce to the ``Brill
wave plus black hole'' distorted black hole data sets described in
Ref.~\cite{Bernstein93b}, and for the Brill wave parameter $q=0$ the
Bowen and York solutions result.

The initial data sets described in Table~\ref{tab:initial} whose name
begins with an {\it r} ({\it r0} through {\it r5}) describe a sequence
of distorted rotating holes of increasing rotation.  The run {\it r0}
has no angular momentum, while run {\it r5} is a pure Bowen and York
black hole without a Brill wave added, but with a significant amount
of rotation.  As we will see in section~\ref{embeddings} below, it is
rotating so fast ($a/m=0.87$) that its horizon cannot be completely
embedded in three dimensional Euclidean space, yet we are able to
evolve it accurately.

Finally we have considered a third family of distorted black hole data
sets that correspond to odd-parity radiation superimposed on
non-rotating black holes.  In this case the three-metric $\gamma_{ab}$
is described by the same Brill wave parameters in Eq.(\ref{brill}) above,
but the extrinsic curvature is taken to be
\begin{mathletters}
\begin{eqnarray}
H_E & = & \Psi^6\, q_G \left(\left(n^\prime+1\right)-
\left(2+n^\prime \right) \sin^2\theta \right) \sin^{n^\prime-3}\theta \\
H_F & = & -\Psi^6\, \partial_\eta q_G
\cos\theta \sin^{n^\prime-1} \theta. \label{odd parity solution}
\end{eqnarray}
\end{mathletters} \\
The parameter $n^\prime$ is used to describe an ``odd-parity''
distortion.  It must be odd, and have a value of at least 3.  This
family of data sets does not describe rotating black holes, as
discussed in Paper~I and in more detail in Paper~III.  The two data
sets in Table~\ref{tab:initial} whose name begins with {\it o} ($o1$
and $o2$) represent a family of odd-parity distortions of
non-rotating holes.  As we will see in section~\ref{embeddings} below,
the angular index $n^\prime$ is related to the dominant $\ell$-mode in
the distortion of these holes.  Both an $n^\prime=3$ and an
$n^\prime=5$ data set are given because the former emitted virtually
all of its radiation in the $\ell=3$ mode and we were also interested
in examining significant $\ell=5$ radiation from an odd-parity
distortion.

We summarize the dimensionless angular momentum parameter ($a/m$) of
these spacetimes in Table~\ref{tab:a/m} where three different
measurements are tabulated. The meanings of the columns are as
follows: $(a/m)_{min}=J/M_{ADM}^2$, and would be the rotation
parameter of the system if none of the radiation escapes to infinity.
$(a/m)_{max}=J/M_{AH}^2$, where $M_{AH}$ is the mass of the apparent
horizon defined by Eq.(\ref{mc}) below, and would be the rotation parameter
of the final system if the area of the black hole did not increase at all
during the evolution.  These two measurements are made on the initial
data sets without evolving them.  Finally $(a/m)_{cr}$ is the value of
$(a/m)$ obtained by measuring the shape of the apparent horizon surface
during the evolution as discussed below and in Ref.~\cite{Anninos93a}.

\section{Near Zone Physics}
\label{nzp}

In this section, we discuss details of the ``near zone'' evolution of
several black holes, focusing on the apparent horizon. We present an
apparent horizon finder, and discuss the dynamics of the apparent
horizon as a tool to study the physics of the near zone.  Although
extremely important to the numerical evolution, the behavior of the
metric functions themselves generally does not illuminate the physics
of the system, and is not discussed here.  We refer the interested
reader to Paper~I for discussion of metric evolution.

\subsection{Horizons}
\label{horizons}

In this section we examine the properties of the black hole apparent
horizons in the spacetimes evolved here.  The apparent horizon is
defined as the outermost trapped surface.  Because a knowledge of the
future is required to identify the event horizon, the concept of an
apparent horizon is useful in numerical relativity.  The location of
the apparent horizon depends only on information known within a given
time-slice, is always inside the event horizon~\cite{Hawking73a} and
it coincides with the event horizon in a stationary spacetime.  As
shown in Ref.~\cite{Anninos94f}, where studies of perturbed event and
apparent horizons were made, apparent horizons can closely approximate
the event horizons even in dynamic spacetimes, and share many
dynamical features with them.  In this paper we focus on the apparent
horizon, but will treat the event horizon of highly distorted black
holes in a future work.

Mathematically the apparent horizon
condition is expressed by requiring the expansion of all outgoing null
normals to the surface to vanish.  This condition can be written
as~\cite{York89}
\begin{equation}
D_a s^a+K_{ab}s^a s^b-trK=0
\label{Trapped Surface Equation}
\end{equation}
where $s^a$ is the outward pointing normal vector to the surface in
the three dimensional subspace.  If such a surface exists, it is said
to be a marginally trapped surface. In some cases multiple surfaces
satisfying this condition can be found, and the outermost one is then
defined to be the apparent horizon.  The apparent horizon finder
searches for a solution to this equation by making an initial guess,
and then iteratively solving the above nonlinear equation via the
Newton-Raphson method until we obtain the desired
accuracy~\cite{Anninos93a}.

The initial guess that we use is an $\eta=$constant surface where the
value of $\eta$ used is the outermost value which satisfies the
condition
\begin{equation}
0=\frac{1}{\Psi^2 \sqrt{A}} \left( 4 \frac{\partial_\eta \Psi}{\Psi}+
\frac{\partial_\eta D}{D} \right)-2 \frac{H_D}{D}
\end{equation}
on the equator ($\theta=\pi/2$).  This may be interpreted as the
condition that the light rays confined to the equatorial plane have
null-expansion.  The radius thus determined provides us with a good
estimate of the position of the apparent horizon.  This technique
proved valuable because it made it easier to follow the horizon when
it jumped outward from the throat at about $t \approx 10 M$.  This
particular complication is a result of using an antisymmetric lapse
that vanishes on the throat of the black hole, preventing evolution
there (except through the shift terms).  The initial apparent horizon
is usually located on the throat (the throat is guaranteed to be a
minimal surface, but not necessarily an apparent horizon). With a
lapse that vanishes on the throat, the horizon often remains there
until some evolution has occurred, at which point it may jump out
discontinuously.  This technique does not require the horizon position
from previous time steps to be recorded, allowing us to find the
horizon as infrequently as we wish, or to find it again if it becomes
lost.  This trick has been quite successful and uses up little CPU
time.  Once we have located the horizon with this method, it can be
analyzed to provide physical information as we describe in the
following sections.

\subsubsection{Masses}
In addition to the ADM mass, which was calculated on the initial
slice, we are interested in obtaining a dynamic mass during the
evolution. The measure we will use for this is an ``areal mass'' based
on the irreducible mass (as given by the apparent horizon) and the
angular momentum of the black hole.

The horizon finder returns the location, $\eta(\theta)$ of the
apparent horizon on a given slice.  In order to study the
geometric properties of the horizon, from which important physical
information can be obtained, it is useful to determine the two
dimensional sub-metric induced on the surface by defining a new
coordinate $\xi$ by the equation $\eta=\eta\left(\theta\right) \xi$.
We now have
\begin{equation}
dl^2=\Psi^4 \left(A d\xi^2+\left(B+A \xi^2
\left(\frac{d\eta}{d\theta}
{\left(\theta\right)}
 \right)^2 \right)d\theta^2+D d\phi^2+2 F d\phi d\theta \right).
\end{equation}
The sub-metric that we are interested in may be found by
letting $\xi$ be equal to 1.  The surface area of the horizon is,
therefore,
\begin{equation}
A=2\pi \int_0^\pi \sin\theta \Psi^4 d\theta
\sqrt{
	\left(
		B+A \left(
			\frac{d\eta}
			{d\theta}\left(\theta\right)
		\right)^2
	\right)D-F^2
}
\end{equation}
For a Schwarzschild black hole the mass is given by $M_{ir}$ below,
and for a Kerr black hole the mass is determined by both the surface
area and angular momentum of the hole~\cite{Christodoulou70}, according to the
relationships
\begin{mathletters}
\begin{equation}
M_{ir} = \sqrt{\frac{A}{16 \pi}},
\end{equation}
and
\begin{equation}
M_{AH}^2 = M_{ir}^2+\frac{J^2}{4 M_{ir}^2}.
\end{equation} \label{mc}
\end{mathletters} \\
Although our spacetimes are not stationary and hence the apparent
horizon does not coincide exactly with the event horizon, these are
quite useful quantities to take for the dynamical mass.  The above
formulations describe the minimum mass of the final Kerr black hole
that is possible at the end of the evolution~\cite{ftn:except}
However, as shown in
Refs.~\cite{Anninos94f,Masso94a} the event and apparent horizons are
often quite close, even in dynamic spacetimes, and the mass described
in Eqs.~(\ref{mc}) should often be a good approximation to the event
horizon based mass.  The masses computed in this way will be used
extensively in analyzing the evolutions presented in
section~\ref{Results} below.

\subsubsection{Oscillations}
\label{oscillation}

Once we have the 2-metric on the horizon's surface we can measure the
horizon geometry.  An example of such a measurement is $C_r$, the
ratio of polar circumference ($C_p$) to the equatorial circumference
($C_e$) of the apparent horizon.  The horizon of a stationary black
hole has a characteristic shape that depends upon its rotation
parameter $a/m$~\cite{Smarr73b}.  For a stationary or a dynamic black
hole, this shape can be used to obtain the rotation parameter, as
discussed in Ref.~\cite{Anninos93a}. There it was shown that the
horizon oscillates about this equilibrium stationary shape with the
quasinormal mode frequencies of the black hole.  We can plot the
shape parameter $C_r$ as a function of time to measure these horizon
dynamics.  Here we extend this technique and apply it to new
spacetimes not considered in Ref.~\cite{Anninos93a}.

To obtain the polar and equatorial circumferences we first transform the
two-metric so that it becomes diagonal.  This is accomplished by
introducing a new coordinate, $\chi$.  The metric on the two surface
defined by the apparent horizon now becomes:
\begin{mathletters}
\begin{eqnarray}
d\chi & = & d\phi-\frac{F}{D \sin\theta}d\theta \\
B' & = & \Psi^4 \left( B+A \left(\frac{\partial\eta}
{d\theta} \left(\theta\right)\right)^2+ \frac{F^2}{D} \right) \\
F' & = & 0\\
D' & = & \Psi^4 D.
\end{eqnarray}
\end{mathletters} \\
The polar circumference and the equatorial circumference
may now be defined as:
\begin{mathletters}
\begin{equation}
C_p = \oint d\theta \sqrt{B^\prime},
\end{equation}
and
\begin{equation}
C_e = \oint d\phi \sqrt{D^\prime}.
\end{equation}
\end{mathletters} \\

It was shown in previous work~\cite{Anninos93a} that the $\ell=2$ and
$\ell=4$ quasinormal mode (QNM) frequencies can be seen in the
function $C_r$, which can also be analyzed to obtain the rotational
speed ($a/m$) of the spacetime. This is accomplished by first fitting
$C_r$ to the fundamental $\ell=2$ QNM, the first overtone $\ell=2$
QNM, and a constant offset from unity.  We assume that the
oscillations are dominated by the $\ell=2$ quasinormal
ringing radiation going down the horizon.  As discussed in
Ref.~\cite{Anninos93a}, the ringing radiation is generated at the
``peak of the potential barrier'' located outside the horizon.  Part
of this signal propagates away from the hole and part goes down the
hole, causing the horizon to oscillate as the signal crosses it.  The
offset is used in the fit because a rotating black hole will not be
spherical (a sphere would have $C_r=1$) but should be oblate (in
equilibrium), producing an ``offset'' in $C_r$.  The fitted value of
the constant offset can then be used to obtain the value of $a/m$ for
the surface, since for a Kerr black hole the oblateness is a unique
function of the rotation parameter. (We note that the following
function is an approximation generally accurate to within $2.5\%$ for
$a/m$ as a function of $C_r$: $a/m=\sqrt{1-\left( -1.55+2.55\,
C_r\right)^2}.$)  In Table~\ref{tab:a/m} we show the value of the
rotation parameter $a/m$ for the family of data sets labeled {\it
r0-r5} obtained with this method.

Finally, as the horizon oscillations should be caused by radiation
from all black hole modes excited during the evolution, not just the
$\ell=2$ modes, one can analyze the function $C_r$ for other modes as
well.  For example, the fit to the shape parameter in terms of the
$\ell=2$ QNM expansion functions can be subtracted from the actual
function $C_r$, leaving a residual of higher modes.  In
Ref.~\cite{Anninos93a} we showed how this remaining signal was
dominated by $\ell=4$ QNM's for some spacetimes.

For the odd-parity distorted holes ({\it o1} and {\it o2}, not shown),
the oscillations in $C_r$ are second order in the metric perturbation.
As a result the spectrum of frequencies in the surface vibrations that
result from the $\ell=2$ and $\ell=3$ QNM's will be much larger.
Therefore the oscillations are more complicated than in the previous
cases.  Further details of the surface oscillations of these black holes
will be given in section~\ref{gaussian_curvature}.

In Fig.~\ref{fig:cr,r4} we plot $C_r$ as a function of time for
simulation {\it r4}.  This is a distorted Bowen and York hole rotating
quite rapidly.  The solid line shows the value of $C_r$ extracted from
the horizon.  The dashed line shows the fit to the two lowest $\ell=2$
QNM's, and the straight horizontal dashed line shows the constant
offset that came from the fit.  Note that the fit matches the
oscillation quite well except at very late times, when we expect the
code to be less accurate, particularly near the horizon where metric
functions tend to develop steep gradients~\cite{Seidel92a}.  This
analysis shows that the black hole is oscillating about a Kerr black
hole with a rotation parameter of $a/m=0.70$, consistent with
estimates made on the initial slice.

In Fig.~\ref{fig:cr,r5} shows a similar plot for the function $C_r$
and its fit for run {\it r5}. Run {\it r5} differs from runs {\it
r0}-{\it r4} in that {\it r5} is a pure Bowen and York black hole.  It
is thus much more noticeably oblate, with $C_r \approx 0.81$, as
opposed to $C_r \approx 0.89$ for the run labeled {\it r4}.  In this
case the analysis shows the rotation parameter to be $a/m=0.87$.

Table~\ref{tab:a/m} gives the results of extracting $a/m$ by this
method in column 3 for runs {\it r0}-{\it r5}.  This table shows that
in all cases the value extracted from $C_r$ during the run is very
nearly the lower limit, $J/M_{ADM}^2$.  This means that in the runs
under consideration in this paper nearly all the radiation is going
into the hole, because the hole achieves nearly the maximum possible
mass (or minimum possible rotation parameter).

\subsubsection{Embeddings}
\label{embeddings}
A valuable feature of the horizon geometry is its three-space
embedding.  By embedding the horizon we can obtain a visual impression
of its overall distortion in a coordinate independent way.  The Kerr
geometry is a sphere in Boyer-Lindquist coordinates, but it is not so
when embedded in three-space.  Instead it is an oblate spheroid that
becomes increasingly flattened along its axis of rotation as its spin
increases. If its rotation parameter $a/m$ exceeds $\sqrt{3}/2$ it is
impossible to completely embed the figure in flat space, as shown by
Smarr~\cite{Smarr73b}.

The procedure for finding the embedding starts by defining a new
coordinate $z$ which is part of a flat 3-metric and is identified
with the 2-metric of our surface:
\begin{mathletters}
\begin{equation}
dz^2+d\rho^2+\rho^2 d\phi^2 = B^\prime d\theta^2+
D^\prime d\phi^2.
\end{equation}
Solving this expression for $z$ gives
\begin{equation}
z=\int d\theta \sqrt{
B^\prime-\left( \partial_\theta \sqrt{
D^\prime} \right)^2 }. \label{z equation}
\end{equation}
and
\begin{equation}
\rho=\sqrt{D}.
\end{equation}
\end{mathletters} \\

This equation is then integrated numerically to obtain the embedding
functions $z(\theta)$ and $\rho(\theta)$.  In Fig.~\ref{fig:r4_embed}
we show the embedding of the evolving horizon for run {\it r4}.  At
time $t=5$ the horizon is still frozen and has its initial prolate
shape with $C_r = 1.228$.  At time $t= 10 M$ the horizon has become
noticeably oblate, overshooting its equilibrium value, and by the late
time of $t=50M$ it has settled down to very near its equilibrium Kerr
value.  The dark solid line shows the shape of the Kerr hole with the
rotation parameter determined from the procedure above.  (The area has
been normalized to take account of the difficulty of resolving the
peak in $A$ that can cause the area to grow spuriously, as discussed
below and in Ref.~\cite{Anninos93a}.)

For a Kerr black hole, when the rotation parameter $a/m$ becomes large
enough and the integrand of Eq.(\ref{z equation}) becomes imaginary,
the embedding ceases to exist.  However, this generally happens only
over a small portion of the horizon surface, near the symmetry
axis. This embedding failure will begin to occur when $a/m=\sqrt{3}/2
\approx 0.87 M$.  The angle $\theta$ (measured from the
$z$-axis) at which the embedding should cease to exist can
be found.  The equation for this angle is
\begin{mathletters}
\begin{equation}
\cos^2\theta = \frac{5+3 b}{3 \left(b-1 \right) }-
\frac{4}{3 \left(1-b\right)^{2/3} x^{1/3}}+\frac{2 x^{1/3}}
{3 \left(1-b \right)^{4/3}},
\end{equation}
where
\begin{equation}
x = 17+10 b+3 \sqrt{3 \left(11+12 b+4 b^2 \right)},
\end{equation}
and
\begin{equation}
b = \sqrt{1-\left(a/m\right)^2}.
\end{equation}
\end{mathletters} \\
Note that for the region $\theta \agt .4 \pi$ the embedding never
disappears.

This effect can actually be used to measure the rotation parameter for
rapidly rotating holes.  For a dynamic rotating black hole, as shown
above the horizon geometry will oscillate about the stationary Kerr
shape.  For an extremely fast rotator or highly distorted black hole
the oscillations may distort the horizon so much that the embedding
fails at times during the evolution (or in extreme cases the embedding
may never exist).  In Fig.~\ref{fig:no_embed} we plot the value of
$\theta$ for which the embedding fails as a function of time for the
case {\it r5}.  This is a pure Bowen and York hole with $J=15$,
corresponding to $a/m=0.87$.

The run labeled {\it r5} is the fastest rotating black hole we have
simulated to date, and it pushes the limits of our code, but it
nevertheless produces a good result, in agreement with theory.  The
straight horizontal line represents the embedding limit of the
analytic Kerr solution with the given value of $a/m=0.87$, which was
the rotation parameter that was extracted from the oscillation about
$C_r$ as discussed above.  Note that this black hole horizon is {\it
never} globally embedded in three dimensional Euclidean space,
although the horizon always exists.  Note also that by measuring the
horizon geometry, we can determine the rotation parameter at least two
different ways: by measuring $C_r$ as in Fig.~\ref{fig:cr,r4} (see
discussion above), and by measuring the angle at which the embedding
ceases to exist (although this only works for holes which have $a/m >
\sqrt{3}/2$).  Furthermore, these measurements agree with estimates of
the rotation parameter determined from the initial data alone.

\subsection{Horizon History Diagrams}
\label{gaussian_curvature}

As we have shown in Ref.~\cite{Anninos93a} it is useful to construct a
``horizon history'' embedding diagram, that shows the evolution of the
surface in time.  In order to bring out details of the local curvature
of this surface, we also compute the Gaussian curvature $\kappa$ of
the horizon surface and map it to a colormap or greyscale.  For
details of this construction we refer the reader to
Ref~\cite{Anninos93a}, but for reference we present the formula
for Kerr here:
\begin{equation}
\kappa=
{{8\,\left( -1 + 5\,b - 3\,\cos (2\,\theta) +
       3\,b\,\cos (2\,\theta) \right) }\over
   {\left( 1 + b \right) \,{m^2}\,
     {{\left( 3 + b + \cos (2\,\theta) -
          b\,\cos (2\,\theta) \right) }^3}}},
\end{equation}
where $b=\sqrt{1-a^2/m^2}$ as before.  The mapping of the Gaussian
curvature onto the surface helps to bring out small deviations in the
local curvature of the surface that would not be apparent in the
embedding diagram itself.

In Fig.~\ref{fig:gauss_odd1} we show a horizon history embedding
diagram for the case {\it o1}, which is an odd-parity distorted
non-rotating black hole.  As we discuss in detail in
section~\ref{Results}, this data set has predominantly $\ell=3$
radiation ($99\%$ of the total energy radiated is carried in this
mode).  As in the cases shown in Ref.~\cite{Anninos93a}, each
$\ell$-mode pattern is qualitatively different from the ones studied
previously.  In Fig.~\ref{fig:gauss_odd2} we show a similar diagram
for the run labeled {\it o2}, which has a significant $\ell=5$
component.  Again, the pattern is qualitatively different from the
$\ell=3$ (or $\ell=2$ or $\ell=4$) pattern, and has not been seen
before.

These distinctive patterns can be understood by analyzing the
expression for the gaussian curvature.  The metric variable $F$ enters
the Gaussian curvature as a second order term (to first order it
vanishes identically).  This function carries the odd-parity radiation
in our gauge, as discussed below.  Therefore if $F$ oscillates
predominantly at the $\ell=3$ normal mode frequency denoted by
$\omega_{\ell=3}$, then the period of oscillation one will see in the
Gaussian curvature plot will not be $\omega_{\ell=3}$ due to this
nonlinearity.  Instead one will see $2 \omega_{\ell=3}$, so the
pattern cycle will repeat twice as quickly as it would have if $F$ had
entered the Gaussian curvature to linear order.  This is what one sees
in Fig.~\ref{fig:gauss_odd1}.  The pattern repeats every $t \approx
5.24 M$ instead of every $t \approx 10.48 M$, which is the period of
the $\ell=3$ fundamental mode.

In Fig.~\ref{fig:gauss_odd2} the $\ell=5$ modes are present to a much
stronger degree, so we see a mixture of the $\ell=3$, and $\ell=5$
modes.  This creates a diamond pattern reminiscent of the diamond
pattern seen on the surfaces of even-parity black hole horizons
generated with an $n=4$ perturbation, as described in
Ref~\cite{Anninos93c}.  In general, however, this pattern is more
complicated than the $n=4$ pattern, as it should include four
frequences: $2 \omega_{\ell=3}$, $\omega_{\ell=3} \pm
\omega_{\ell=5}$, and $2 \omega_{\ell=5}$.

\section{Far Zone Physics}
\label{farzone}
\subsection{Wave Mode Extraction}

In previous sections we have analyzed properties of the near zone
features of these new black hole data sets, focusing on geometric
measures of the horizon.  These effects provide important probes of
the physics of the near zone, and can also be used to determine
important properties of evolving black holes that may not be known,
but for the most part these effects are not measurable by gravitational
wave experiments such as LIGO.  In this section we turn to analysis of
the physics of the far zone, where important features such as
gravitational waves can actually be observed.

One of the principal features of a dynamical black hole spacetime
which has been possible to study analytically is linearized
gravitational waves.  Therefore, it is useful for us to measure them
in highly distorted black hole spacetimes to determine information
that linear theory cannot predict, such as the final mass of a black
hole in a perturbed spacetime, or the waveform emitted during the
nonlinear generation of gravitational waves, and to provide several
useful checks of our code.  Although the processes that generate these
waves may be highly nonlinear, far from the hole the waves may be
treated as linear perturbations on a fixed background.  For all the
simulations presented in this paper we extracted the radiation at a
distance $r=15 M$.

The radiation energy in a black hole spacetime is described by certain
gauge-independent variables constructed from the metric and its
derivatives.  The technique was originally developed by Abrahams and
Evans~\cite{Abrahams90} and applied to black hole
spacetimes in Ref.~\cite{Abrahams92a}.  There are two classes of
gauge-invariant radiation quantities representing the two degrees of
freedom of the field, even-parity and odd-parity.  Both are present in
our spacetimes and we describe them below.

\subsubsection{Even-parity}

We used the same wave extraction routines for the even-parity wave
forms as used in Ref~\cite{Abrahams92a}.  The routines were developed
to extract waveforms from perturbed {\it Schwarzschild} black hole
spacetimes that oscillate about a spherical background.  Because the
distorted {\it Kerr} metric does not settle down to something that is
spherical, but rather it settles down to something oblate, we should
expect to see this effect in the extraction process, resulting in an
offset in the $\ell=2$ and $\ell=4$ waveforms.  Thus, the wave does
not oscillate about zero, but instead will be offset from zero
depending on both the rotation parameter and the radius at which the
wave is extracted.

However, the level of this offset is small and its variation rather
too sensitive so that we have not been able to use it to measure the
rotation parameter reliably.  As shown below, we do see this effect in
waveforms extracted from distorted rotating black holes, and the
offset value is related to the rotation parameter as expected.  The
offset we obtain from the waveforms is of the right general magnitude
and sign for a given rotation parameter.  In principal one can account
for the fact that the system is a perturbed Kerr black hole, but to
date we have not carried out this analysis. However, as we show below,
the extraction assuming a spherical Schwarzschild background can be
quite useful without modification.

The even-parity wave extraction is given here.
Note that this extraction formula assumes a Schwarzschild background.
For each $\ell$-mode we can extract independent radiation waveforms.
The gauge invariant, even-parity wave function is given by
\begin{mathletters}
\begin{equation}
\psi^{even} = \sqrt{\frac{2(\ell-1)(\ell+2)}{\ell(\ell+1)}}
\frac{\left(4 r S^2 k_2+\ell(\ell+1)r k_1 \right)}{\Lambda},
\end{equation}
and
\begin{equation}
\Lambda = \ell(\ell+1)-2+\frac{6 M}{r},
\end{equation}
where the Moncrief functions~\cite{Moncrief74} are given by
\begin{eqnarray}
k_1 &=& K+S r \frac{\partial G}{\partial r},\\
k_2 &=& \frac{H_2}{2 S}-\frac{1}{2 S^{\frac{1}{2}}} \frac{\partial}{\partial r}
\left( r S^{-\frac{1}{2}} K \right),
\end{eqnarray}
with
\begin{eqnarray}
S &=& 1-\frac{2 M}{r}, \\
r &=& e^{\eta},
\end{eqnarray}
and finally the Regge-Wheeler~\cite{Regge57} perturbation functions are
defined in terms of the 3-metric via
\begin{eqnarray}
H_2 &=& \frac{2 \pi}{\hat{A}^2} \int^\pi_0 \Psi^4 A\, Y_{\ell 0}\sin\theta
d\theta, \\
\hat{A}^2 &=& \frac{1}{2} \int_0^\pi \Psi^4 A \sin\theta\, d\theta, \\
G &=& \frac{2 \pi}{R^2} \int_0^\pi \Psi^4 \frac{
\left(B-D\right) \left(-\cos\theta\, Y_{l0,\theta}+\sin\theta\, Y_{l0,\theta},
\right)d\theta
}{ \ell \left( \ell+1 \right) \left( \ell+2 \right) \left( \ell-1 \right) }, \\
K &=& \frac{\ell \left( \ell+1 \right)}{2} G+ \frac{\pi}{R^2}
\int_0^\pi \Psi^4 \left(B+D \right)\sin\theta\, Y_{\ell 0} d\theta, \\
R^2 &=& \frac{1}{2} \int_0^\pi \Psi^4\,B\,\sin\theta\,d\theta.
\end{eqnarray}
\end{mathletters} \\
With the normalization of $\psi$ given above one can show that the
total radiated energy in each $\ell$-mode is given by
\begin{equation}
E=\frac{1}{32 \pi} \int dt \left( \partial_t \psi \right)^2.
\end{equation}
Complete details of this extraction procedure are provided in
Ref.~\cite{Abrahams92a}.

\subsubsection{Odd-parity}
\label{odd-parity}

In terms of the Regge and Wheeler formalism~\cite{Regge57} there are
various ways to construct the odd-parity gauge-invariant variable that
measures the radiation in the system~\cite{Moncrief74}.  These
measures are not linearly independent, and each is constructed from
two of the variables $E$, $F$, and $\beta^\phi$, which correspond to
Regge and Wheeler's variables $h_1$, $h_2$, and $h_0$ after
appropriate angular integrals have been performed.  The
gauge-dependent Regge-Wheeler variables can be extracted from the
metric as
\begin{mathletters}
\begin{eqnarray}
h_0 &=& \sqrt{\frac{(\ell+1)!}{(\ell-1)!}}
\oint d\Omega\,  \beta_\phi\, \partial_\theta Y_{\ell 0},\\
h_1 &=&  \sqrt{\frac{(\ell+1)!}{(\ell-1)!}}
\oint d\Omega\,  \Psi^4 E\, \partial_\theta Y_{\ell 0},
\end{eqnarray}
and
\begin{equation}
h_2 = \frac{1}{2} \sqrt{\frac{(\ell+2)!}{(\ell-2)!}}
\oint d\Omega\,\Psi^4 F \left(\partial_\theta-\cot\theta
\right) \partial_\theta Y_{\ell 0}.
\end{equation}
\end{mathletters} \\
{}From these perturbation functions one can construct the following
gauge-invariant quantities, two of which are given in
Ref~\cite{Cunningham78}:
\begin{mathletters}
\begin{eqnarray}
\psi_{02} &=& h_0+\frac{1}{2} \partial_t h_2 \label{extracta},\\
\psi_{01} &=& r^2 \partial_r \left( \frac{h_0}{r^2} \right) -
\partial_t h_1,
\end{eqnarray}
and
\begin{equation}
\psi_{12} = h_1+\frac{1}{2} r^2 \partial_r \left(
\frac{h_2}{r^2} \right)\label{extractc}.
\end{equation}
\end{mathletters} \\

In our code we have used our gauge freedom to eliminate the metric
function $E$ ($h_1$) and the method we use to extract the radiation is
effectively Eq.(\ref{extractc}).  It is also possible to measure the
odd-parity radiation through Eq.(\ref{extracta}) as we show below.
Note that for the odd-parity case, there is no ``non-spherical
odd-parity'' part in a Kerr black hole (in $\psi_{12}$) so there is no
offset expected nor observed for these waveforms.

The particular energy integrals for the odd parity modes are given here.  Note
that these expressions are specific to our gauge and a flat space background
metric.  We first normalize our gauge invariant odd-parity wave function,
based on Eq.(\ref{extractc}) as follows:
\begin{mathletters}
\begin{equation}
\psi^{odd}_{\ell=3}=\int_0^\pi d\theta\, \frac{1}{2} \sqrt{105 \pi}
\cos\theta \sin^3\theta \partial_\eta F
\end{equation}
and
\begin{equation}
\psi^{odd}_{\ell=5}=\int_0^\pi d\theta \frac{1}{16} \sqrt{1155 \pi}\left(
5 \cos\theta+3 \cos\left( 3\theta\right) \right) \sin^3\theta \partial_\eta F.
\end{equation}
\end{mathletters} \\
For general $\ell$ the energy expression is this
\begin{equation}
\psi^{odd}=\int_0^\pi 2 \pi \sqrt{2\frac{(\ell-2)!}{(\ell+2)!}}
d\theta\, \partial_\eta F \sin\theta
\left( \partial_\theta-\cot\theta \right) \partial_\theta Y_{\ell 0}.
\end{equation}

In principle one can extract arbitrarily high $\ell$-modes.  In
practice we have only examined the $\ell=3$ and $\ell=5$ modes to
date.  With these normalizations, the energy radiated by each
$\ell$-mode is given by
\begin{equation}
E=\frac{1}{32 \pi} \int dt \psi^2,
\end{equation}
as in Ref.~\cite{Cunningham78}.

\subsection{Measurement of Frequencies}
\label{fourier}

To analyze quantitatively the frequency spectrum of our radiation we
used the following Fourier technique.  Because the quasinormal modes
of black holes are all damped the Fourier spectrum of their
frequencies will be spread out.  To counter this effect, before
analyzing a waveform we premultiply it by a factor $\exp(\lambda t)$,
where $\lambda$ is chosen to approximately cancel the effects of the
damping (see Ref~\cite{Bretthorst88} for a discussion of such
techniques).  As it happens, the damping time for the fundamental mode
is given approximately by $\lambda \approx .09/M$, relatively independent of
the $\ell$-value, so this technique is helpful for all modes.

We next decompose the frequency spectrum using the procedure described
below.  The technique used here to analyze the waveforms is preferable
to an FFT, since an FFT may provide poor frequency resolution for the
wavelengths under study unless the sampling interval ($\Delta T$) is
quite long.  For example, to distinguish between the peaks in $\ell=2$
and $\ell=3$ QNM's would require a ``sampling period'' greater than
$\Delta T \approx 60 M$ of oscillations.  The Fourier technique we
present below, combined with the premultiplication technique, can
distinguish between them with only about $\Delta T \approx 17 M$ of
oscillations.  We approximate the Fourier transform function and its
inverse by
\begin{mathletters}
\begin{eqnarray}
\hat{f}(\omega) &=& \frac{1}{\sqrt{2 \pi}}
\int_{-\infty}^\infty f(t) e^{-i \omega t } dt \approx
\sum_i f_i \hat{\delta}_i(\omega) \Delta t\\
f(t) &=& \frac{1}{\sqrt{2 \pi}}
\int_{-\infty}^\infty \hat{f}(\omega) e^{i \omega t} d\omega \approx
\sum_i f_i \delta(t-t_i) \Delta t,
\end{eqnarray}
where we have defined
\begin{equation}
\delta(t-t_i) =
\frac{1}{\sigma \sqrt{\pi}} e^{-(t-t_i)^2/\sigma^2},
\end{equation}
and
\begin{equation}
\hat{\delta}_i(\omega) = e^{-i\omega t_i-\sigma^2 \omega^2/4}.
\end{equation}
\end{mathletters} \\
In the above equations and $f(t)$ is the Regge-Wheeler or Zerilli
function, $(f_i,t_i)$ are the set of points produced by the numerical
code (we should find that $f_i \approx f(t_i)$), and $t$ is the time.  The
value of $\sigma$ was set to $\Delta t$.

Note that the presence of the first overtone of a quasinormal mode
frequency cannot be as easily detected with this technique (since the
damping is so much greater).  It manifests as a small shift in the
position of the peak of the fundamental mode.  The position of the
peak will also be affected by the artificial growth in the mass that
occurs at late times, which is manifested in the waveform by
lengthening of the wavelength of the radiation as discussed in
Ref.~\cite{Abrahams92a}.  This artificial lengthening will
effectively add a mixture of other frequencies to the spectrum,
further broadening the peak.
\section{Results and Discussion}
\label{Results}

In this section we discuss results from evolutions of the initial data
sets listed in Tables~\ref{tab:the_k's}
and~\ref{tab:initial},
combining many of the analysis techniques discussed in the previous
sections.  Together these tools provide a thorough and remarkably
consistent physical picture of the evolution of these black hole
spacetimes.

\subsection{Odd-parity Distorted Schwarzschild Black Holes}
\label{odd}
In this section we give results for evolution of the new class of data
sets we constructed and discussed briefly in Ref.~\cite{Brandt94b}.
These evolutions correspond to runs labeled {\it o1} and {\it o2}.
These data sets do not possess angular momentum, but they do possess
odd-parity radiation, in contrast to the ``Brill wave plus black hole
spacetimes'' discussed previously in Ref.~\cite{Bernstein93b}.
Strictly speaking, they are odd-parity distorted Schwarzschild black
holes, and not rotating holes.  Nevertheless, the odd-parity
distortions give rise to ``rotation-like'' features.  We note that
these black hole data sets also contain even-parity radiation,
although of a much lower amplitude.

The run labeled {\it o1} represents a spacetime with virtually only an
$\ell=3$ distortion, as almost all the energy is radiated in that
mode.  Fig.~\ref{fig:l=3,o1} shows the $\ell=3$ waveform extracted by
the gauge-invariant waveform extraction method described in
section~\ref{odd-parity} above. As one can see, it is almost
impossible to distinguish the fit from the data.  As we show in
Table~III, $99.96\%$ of the total energy radiated is carried by the
$\ell=3$ mode. In Fig.~\ref{fig:l=5,o1} we show the $\ell=5$ waveform
extraction for the same run.  It is interesting to note that even
though it does not contribute significantly to the energy, and the
signal amplitude is nearly four orders of magnitude smaller than the
$\ell=3$ signal, the $\ell=5$ signal is still easily fit to the proper
mode.

For run {\it o1} we extracted the $\ell=3$ radiation mode from both
$\beta^\phi$ and $F$.  Both extractions are plotted in
Fig.~\ref{fig:odd_from_shift}.  The solid line illustrates the
extraction from the shift, the dotted line traces the value extracted
from $F$.  The waveform extracted from $\beta^\phi$ was normalized so
that it had the same lower bound as the waveform extracted from $F$.
This shows that it is possible to perform the extraction from the
shift, although we have only done so for this test.

Next we consider the run labeled $o2$, which has a larger $\ell=5$
signal.  The $\ell=5$ extracted mode, shown in Fig.~\ref{fig:l=5,o2},
matches the fit to the quasinormal modes quite well, as in the
previous case.  (The $\ell=3$ waveform, not shown here, matches its
quasinormal mode fit with the same level of accuracy as in the
previous case {\it o1}.)  After measuring the $\ell=5$ frequency from
our extracted waveform via the Fourier transform technique discussed
in section~\ref{fourier} above, we discovered that its quasinormal
modes were not tabulated in the literature.  We computed the real part
of the frequency to be $1.00\pm.01$.  Subsequent calculations by
Edward Leaver, based on black hole perturbation
theory~\cite{Leaver86}, yield the result $1.012$ which agrees with our
result to within $1\%$.  In Fig.~\ref{fig:fourier5} we compare the
analytic $\ell=5$ frequency with the Fourier transform (as implemented
by the technique described above) of our data for the run labeled {\it
o2}.  A dotted line is placed on the graph to show where the peak
should be for a pure $\ell=5$ wave.  The secondary peaks do not
represent real frequencies, rather they are an artifact of the
extraction process.

Finally, we turn to a calculation of the energies and masses in the
system.  The mass of the apparent horizon for the run labeled {\it o1}
is plotted in Fig.~\ref{fig:mass,o1}.  The solid line shows the
instantaneous mass of the horizon, defined by Eq.(\ref{mc}), and the
long dashed line represents the total energy radiated away from the
black hole during the evolution. All energies in this plot are
normalized to the ADM mass, and therefore it can readily be seen that
the sum of the final apparent horizon mass (as measured at $t=25 M$)
and the energy emitted through radiation (as calculated by integrating
the even- and odd-parity radiation functions for $\ell=2,3,4$ and $5$
for the entire time of the run) add up to the ADM mass.  The very
slight dip in the mass at around $30 M$ becomes less noticeable with
increasing resolution.

After about $t=35 M$ the peak developing in the radial metric function
is not adequately resolved.  As the horizon is located near this peak,
as discussed in Paper~I, its area is not accurately computed after
this time.  However, the simulation can be continued until about
$t=100M$ and accurate waveforms can be extracted throughout.
Furthermore, the calculations of the {\it shape} of the horizon
continue to be accurate throughout the evolution, even though the area
becomes inaccurate.

\subsection{Kerr Black Holes}
\label{Kerr}

Although a Kerr black hole is stationary and not dynamical it does
provide a useful test case for evolution.  Furthermore, as it is the
``equilibrium'' black hole about which all other black hole spacetimes
considered in the paper oscillate, it is useful to study its
properties.

Because there is no radiation in the system we do not expect the
horizon to oscillate.  In Paper~I we used Kerr as a test case to show
that our code was able to evolve the Kerr spacetime accurately by
computing its angular momentum during the evolution.  Here we apply
two of the near zone measurements developed in section~\ref{nzp} for
these spacetimes to show the numerical properties of these evolved
black holes.

In Fig.~\ref{fig:k1,mass} we show the mass $M_{AH}$ of an $a/m=.48$
black hole (labeled $k1$) as a function of time for three different
resolutions.  The horizon mass should be strictly conserved and equal
to the ADM mass for a stationary black hole.  The high resolution
($300 \times 30$) mass is plotted as a solid line, the medium
resolution ($150 \times 24$) mass is plotted as a dotted line, and the
low resolution ($75 \times 12$) mass is plotted as a dashed line.  The
mass is nearly constant until the large peaks discussed in Paper~I
develop and cannot be adequately resolved.  By $t=40M$, the error in
the mass is still less than $.1\%$ at all but the lowest resolution
and by $t=60 M$ the error is still only $.5\%$.  Although the apparent
horizon mass does not exactly equal the ADM mass, it is within the
error expected given the level of angular resolution of the horizon.
The jaggedness of the lower resolution lines is a numerical effect.  A
new bump occurs each time the interpolator in the horizon finder
changes the set of grid zones it uses.  It is only visible because of
the small range of $M_{AH}/M_{ADM}$ which we are viewing.

In Fig.~\ref{fig:k2,mass} we show a similar graph for run $k2$,
corresponding to a Kerr black hole with rotation parameter $a/m=.68$.
Despite the higher rotation, all features noted for run {\it k1} are
preserved to a high degree.  Because there is more angular variation
in the metric there is slightly less agreement between $M_{AH}$ and
$M_{ADM}$.  However, the pure Kerr spacetimes do pose more numerical
problems than the distorted spacetimes, as noted in Paper~I.  In run
{\it k2} the axis instability develops sooner at the highest
resolution, and the results are unreliable after about $t=50M$.

\subsection{Distorted Bowen and York}
\label{distortedBY}

In this section we discuss results for the six initial data sets
labeled {\it r0 - r5}.  This represents a sequence of distorted black
holes with increasing values of the rotation parameter, $a/m$,
beginning with a nonrotating case, labeled {\it r0}.

The first case we discuss is run {\it r0}, which has the same
construction as the ``Brill wave plus Black Hole Spacetime'' discussed
in detail in Ref.~\cite{Bernstein93b}.  This spacetime was evolved
with an antisymmetric lapse across the black hole throat, so although
the data set is in the same class as those evolved in
Ref.~\cite{Abrahams92b}, it has been evolved with a new code capable
of handling rotating black holes, and with a different slicing
condition.  The same simulation was discussed in Paper~I, where a
comparison of metric functions was made to the evolution obtained with
the code described in Ref~\cite{Bernstein93b}.  In Fig.~\ref{fig:l=2,r0} and
Fig.~\ref{fig:l=4,r0} we show the now familiar $\ell=2$ and $\ell=4$
waveforms extracted from the evolution, with a fit of the two lowest
quasinormal modes in each case.  The match is excellent.  It is also
important to note that no odd-parity radiation is present in this
system and there are no $\ell=3$ or $\ell=5$ waveforms to show. ( It
is important to point out that odd-parity does $not$ mean odd-$\ell$,
and even-parity does $not$ mean even-$\ell$.  It simply happens that
with equatorial plane symmetry, which is present here, there are no
odd-$\ell$, even parity modes, nor are there even-$\ell$,
odd-parity modes.)  The new dynamical variables used in this evolution
remain exactly at zero when there is neither odd-parity distortion nor
rotation.

In Fig.~\ref{fig:mass,r0} we show the mass of the apparent horizon for
run {\it r0}, as defined in section~\ref{horizons} above.  Note that
initially it does not change, as the antisymmetric lapse freezes the
evolution at the throat where the apparent horizon is found on the
initial data set.  Then at about time $t=10 M$, the horizon is found
out at a larger radius and evolves slowly outward, its mass increases
as more gravitational wave energy crosses the horizon surface into the
hole.  (This ``jumping-out'' of the horizon is a common property of
all simulations we have performed with an anti-symmetric slicing
condition as discussed in section~\ref{horizons} above.)  After a time
of about $t=25M$, there is a slow upward drift in the mass due to
difficulties in resolving the peak in the radial metric function $A$
that develops near the horizon~\cite{Anninos93a}.  By this time,
nearly all of the in-going gravitational wave energy has entered the
black hole, and we compute the final mass of the hole to be $M_{AH} =
.915 M_{ADM}$.  We also have computed the total energy radiated through
the dominant $\ell=2$ mode, as shown in Table~\ref{tab:more_masses},
and plot it in Fig.~\ref{fig:mass,r0}.  Note that all the energy is
accounted for to a high degree of accuracy.  The total energy radiated
by the black hole, as computed by integrating the Zerilli function,
plus the final mass of the black hole, as determined by the mass of
the apparent horizon, gives the total ADM mass of the spacetime, as
one would expect.

Next we turn to the run labeled {\it r1}.  This calculation is similar
to the previous run except that a noticeable amount of energy is now
being radiated in the $\ell=3$ quasinormal mode, as shown in
Fig.~\ref{fig:l=3,r1}.  The value of $a/m$ is extracted from $C_r$,
demonstrating that this technique is effective for small values of
$a/m$, as shown in Fig.~\ref{fig:cr,r1} (although $a/m=.35$ is
considered to be ``small'' in this study, the vast majority of black
holes in the real universe are probably smaller than
this~\cite{Lamb94}).

Both runs {\it r2} and {\it r3} were typical of many simulations. Each
provides waveforms that fit equally as well as {\it r4}, each locks on
to the appropriate limit slice, and in both cases the radiation energy
plus the final apparent horizon mass add up to the ADM mass.  We show
the plots for the radiation in the $\ell=2$, $3$, $4$, and $5$ modes
for run {\it r2} in figures
Figs.~\ref{fig:l=2,r2},
\ref{fig:l=3,r2},
\ref{fig:l=4,r2}, and
\ref{fig:l=5,r2} respectively.
In each case we fit the wave function to the fundamental and first
harmonic mode for the appropriate value of $\ell$.  The value of $a/m$
used in the frequency fits was derived from the value extracted from
the apparent horizon, although it made very little difference whether
the dependence of the QNM frequencies on the rotation parameter was
accounted for.  Except in the case of extreme rotations, the QNM
frequencies depended only weakly on $a/m$~\cite{Leaver86,Seidel90a}.
In the case of the even parity radiation we include a constant offset
in the fit to account for the non-sphericity of the spacetime.  The
fit is good in all cases.

The calculation {\it r4} is the second fastest rotator in this
series.  This rotation parameter ($a/m=.7$) was attained by reducing
the strength of the Brill wave rather than by increasing the value of
the angular momentum, and it is thus a different construction from
runs {\it r0} to {\it r3}.  Large positive amplitude Brill waves make
it difficult to create spacetimes with large values of $a/m$, as they
increase the mass significantly.  We show the plots for the radiation
in the $\ell=2$, $3$, $4$, and $5$ modes for this run in
Figs.~\ref{fig:l=2,r4},
\ref{fig:l=3,r4},
\ref{fig:l=4,r4}, and
\ref{fig:l=5,r4}.
In each case we fit the wave function to the fundamental and first
harmonic mode for the appropriate value of $\ell$.  Again, the fit is
good in all cases.

The run labeled {\it r5} is a special case.  No ``Brill wave''
distortion is present in the spacetime except the distortion necessary
to make the hole conformally flat.  Both the 3-metric and the
extrinsic curvature are given by the simple solution of Bowen and
York: $H_E = 3 J \Psi^6$ and $H_F = 0$.  Bowen and York black holes
are tricky to evolve, because they are so close to the Kerr black
hole, yet this is the only method by which we have obtained a hole
that rotates so rapidly.  The presence of a Brill wave both stabilizes
the spacetime and reduces the rotation parameter.

We note that in all these spacetimes, as the rotation is increased, an
ever smaller fraction of the energy available as radiation is able to
escape from the black hole.  The dominant energy source in this
problem is the imposed gravitational wave at $\eta=1.0.$ Because of
our coordinate transformation a constant $\eta$ value maps to an ever
decreasing value of $r$ as $a/m$ approaches unity, and the width of the
packet in the $\eta$ direction is effectively decreasing as $a/m$ is
increased, therefore less energy is distributed in the region outside
the horizon.

We note that in all these spacetimes it is approximately true that
$M_{rad}+M_{AH,final}$ is equal to $M_{ADM}$.  The ``final'' apparent
horizon mass is, however, really the mass as given around $30-40 M$,
since after that time large gradients in the radial metric variable
$A$ make it difficult to accurately obtain the horizon and bring about
the well-known artificial inflation of the hole's mass at late times.

\section{Conclusions and Future Work}
\label{conc}

Rotating vacuum black hole spacetimes are a significantly more complex
type of axisymmetric vacuum spacetime than their non-rotating
counterparts.  In this paper we have applied a new code, designed to
evolve rotating black holes, to evolve a new family of rotating black
hole initial data sets.  We developed a series of tools to analyze
these spacetimes, including even- and odd-parity waveform extraction
and various techniques to study the apparent horizons of these
spacetimes.  We showed that measurements of the horizon can be used to
determine the mass, angular momentum, and oscillation frequency of the
black holes.  Distorted rotating holes are shown to oscillate about
their stationary ``Kerr'' equilibrium configurations.

Studies of a new class of odd-parity distorted non-rotating black
holes were also made, and the waveforms and horizons were analyzed in
detail.  In all cases the normal modes of the black holes were shown
to be excited and dominate the wave forms.  We also determined, for
the first time, the $\ell=5$ quasinormal modes of a black hole by
direct measurement of our numerically evolved spacetimes which were
later verified by black hole perturbation theory techniques by Leaver.
Finally, we were able to determine the final mass of the black hole by
measuring the horizon area and compared it to the total ADM mass of
the spacetime.  The difference between these measurements agreed with
the total energy radiated to within a few percent, showing the high
degree of accuracy we are able to obtain in these studies.

There is more work to be done in this for which our code will be
useful.  We plan to discuss, in future papers, other features of the
initial data sets and to look at new data sets.  We plan to study
spacetimes with ``cosmic-screw'' type symmetry, which allows for
rotation with equal and opposite amounts above and below the
equatorial plane.  Ultimately, we plan to study the collision of
rotating black holes.

More information about our group and the research we do is available
through a world wide web server maintained by our group.  We plan to
have movies of some of the simulations discussed in this paper
available there.  The URL for our group is
http://jean-luc.ncsa.uiuc.edu.

%
%
%
%
%
%
%

\section{Acknowledgements}
We would like to thank Andrew Abrahams, Pete Anninos, David Bernstein,
Larry Bretthorst, Karen Camarda, Greg Cook, David Drabold, David
Hobill, Peter Leppik, Larry Smarr, and Wai-Mo Suen for many helpful
suggestions throughout the course of this work.  We are also
especially grateful to Edward Leaver for computing the $\ell=5$
quasinormal modes of black holes so that we could compare them to our
results.  This work was supported by NCSA, and by grants NSF
PHY94-07882 and ASC/PHY93-18152 (ARPA supplemented).  The calculations were
performed at
NCSA on the Cray Y-MP and at the Pittsburgh Supercomputing Center on
the Cray C-90.


\begin{thebibliography}{10}
\bibitem{ftn:except}
Note that if $J/M_{ir}^2 > 2$ then as the horizon area increases $M_{AH}$ will
  actually decrease. However, no Kerr solution has this perverse value of
  $J/M_{ir}^2$, and the final solution we evolve towards must be a Kerr
  solution. In an axisymmetric spacetime this means, therefore, that during the
  subsequent evolution the hole's area must increase until it is at least
  $M_{AH}=\sqrt{J}$. This extreme case does not occur in any spacetime
  considered here.
\end{thebibliography}


\begin{figure}
\caption{In this figure we plot the function $C_r$ (the ratio of polar
to equatorial circumference of the apparent horizon) as computed by
our code for run labeled {\it r4}, a distorted rotating black hole.  We have
removed the early part of this plot where the horizon ``jumps out'' so
that we can more clearly see the surface oscillations.  The line
labeled ``equilibrium'' is the offset from sphericity determined by
our fit.  This offset implies the system is oscillating about an
``equilibrium'' Kerr black hole with a rotation parameter of $a/m=.70$
black hole. }
\label{fig:cr,r4}
\end{figure}

\begin{figure}
\caption{This figure shows $C_r$ (the ratio of polar to equatorial
circumference of the apparent horizon) as a solid line and an $\ell=2$
quasinormal mode fit form run {\it r5}.  The horizontal short and long
dashed line is the offset from sphericity, determined by the fit,
showing that the black hole is oscillating about a Kerr hole with
rotation parameter $a/m=0.87$.}
\label{fig:cr,r5}
\end{figure}

\begin{figure}
\caption{
This figure shows the 3-space embedding of the distorted rotating
black hole run labeled {\it r4} at various times.  Although the black
hole's apparent horizon is initially prolate, it eventually settles down to
the appropriate oblate Kerr black hole shape denoted by a dark solid line.}
\label{fig:r4_embed}
\end{figure}

\begin{figure}
\caption{This plot shows the position at which the Euclidean
3-space embedding of the apparent horizon vanishes for run labeled
{\it r5}, a Bowen and York rotating black hole, with angular momentum
parameter $J=15$ (solid line).  As discussed in the text, for rapidly
rotating holes the horizon embedding fails near the axis at an angle
determined by the rotation parameter.  The dashed line shows the angle
at which the embedding fails for an ``equilibrium'' Kerr black hole
with $a/m=0.87$.}
\label{fig:no_embed}
\end{figure}

\begin{figure}
\caption{
We show the horizon history diagram for the run labeled {\it o1}.  The
Gaussian curvature $\kappa$ is mapped to a grayscale on the surface of
the horizon.  The vertical axis gives the angular location on the
horizon, and the horizontal axis traces out the time development.  As
discussed in the text, we see a pattern that repeats at twice the
$\ell=3$ QNM frequency in accord with theoretical expectations.  In
the figure, dark regions are more highly curved than light regions.}
\label{fig:gauss_odd1}
\end{figure}

\begin{figure}
\caption{
In this plot we show the horizon history diagram for the run labeled
{\it o2}.  Hence the Gaussian curvature shows a diamond pattern, a
result of the mixing of several frequencies of the $\ell=3$ and
$\ell=5$ quasinormal modes.  In the figure, dark regions are more
highly curved than light regions.  See text for details.}
\label{fig:gauss_odd2}
\end{figure}

\begin{figure}
\caption{We plot the extraction of the $\ell=3$ waveform from
the odd-parity distorted non-rotating hole, run labeled {\it o1}.
The dashed line shows the fit of this waveform to the two lowest $\ell=3$
quasinormal modes.  Practically
all of the radiated energy is in this mode.}
\label{fig:l=3,o1}
\end{figure}

\begin{figure}
\caption{This figure shows the numerically extracted $\ell=5$
waveform and its fit to the two lowest $\ell=5$ quasinormal modes for
the run labeled {\it o1}.}
\label{fig:l=5,o1}
\end{figure}

\begin{figure}
\caption{In this figure we compare the $\ell=3$ wave form extracted
from the $\phi$ shift $\beta^\phi$ (dashed line) and from the metric
variable $F=g_{\theta\phi}/\Psi^4$ (solid line), where $\Psi$ is the
conformal factor.  Although we generally use the
extraction based on the 3-metric, this demonstrates that other
techniques can be used.  Note that the waveform extracted from the
$\beta^\phi$ was normalized to have the same lower bound as the
shift extracted from $F$.}
\label{fig:odd_from_shift}
\end{figure}

\begin{figure}
\caption{We show the extraction of the $\ell=5$ waveform from
an odd-parity distorted non-rotating hole, run labeled {\it o2}.  It is clear
that the fit (dashed line) agrees well with the numerically extracted
function (solid line).}
\label{fig:l=5,o2}
\end{figure}

\begin{figure}
\caption{This figure shows a Fourier transform of the data extracted
from the run labeled {\it o2}.  The dotted line shows real part of the
$\ell=5$ frequency, computed via black hole perturbation theory.}
\label{fig:fourier5}
\end{figure}

\begin{figure}
\caption{
This figure shows the apparent horizon mass defined in the text, for
the run labeled {\it o1} normalized by the ADM mass (solid line).
After about $t=40 M$ the area of the horizon slowly drifts up due to
difficulties in resolving metric functions, causing an over-estimation
of the horizon mass. (See text.) The short-dashed line is drawn at $1$
representing the total mass of the spacetime.  The gap between the
short-dashed and long-dashed lines is the amount of energy emitted
through radiation.  As one can see, the mass of the horizon
and the energy of the radiation add up to the ADM mass.}
\label{fig:mass,o1}
\end{figure}

\begin{figure}
\caption{This figure shows the apparent horizon mass for run
{\it k1} ($a/m=0.48$) at three resolutions.  It is clear that the
apparent horizon mass becomes more constant and agrees better with the
ADM mass as resolution is increased.  At all but the lowest resolution
we have about $0.5\%$ accuracy in the apparent horizon mass at the
late times shown.}
\label{fig:k1,mass}
\end{figure}

\begin{figure}
\caption{This figure shows the apparent horizon mass for run
{\it k2} ($a/m=0.68$) at three resolutions.  It is clear that the
apparent horizon mass becomes more constant and agrees better with the
ADM mass as resolution is increased.  At all but the lowest resolution
we have about $0.1\%$ accuracy in the apparent horizon mass.}
\label{fig:k2,mass}
\end{figure}

\begin{figure}
\caption{
This figure shows the numerically extracted $\ell=2$ waveform (solid
line) and the least squares fit to the two lowest $\ell=2$ quasinormal
mode for this distorted non-rotating black hole.  This run was
labeled {\it r0}.}
\label{fig:l=2,r0}
\end{figure}

\begin{figure}
\caption{This figure shows the numerically extracted $\ell=4$
waveform and its fit to the two lowest $\ell=4$ quasinormal modes for
the run labeled {\it r0}.}
\label{fig:l=4,r0}
\end{figure}

\begin{figure}
\caption{
This figure shows the apparent horizon mass defined in the text, for
the run labeled {\it r0} normalized by the ADM mass (solid line). The
short-dashed line is drawn at $1$ representing the total mass of the
spacetime.  The gap between the short-dashed and long-dashed lines is
the quantity of radiation emitted.  The mass of the horizon and the
energy of the radiation add up to the ADM mass.}
\label{fig:mass,r0}
\end{figure}

\begin{figure}
\caption{This figure shows the numerically extracted $\ell=3$
waveform and its fit to the two lowest $\ell=3$ quasinormal modes for
the run labeled {\it r1}.}
\label{fig:l=3,r1}
\end{figure}

\begin{figure}
\caption{This figure shows $C_r$ (the ratio of polar to equatorial
circumference of the apparent horizon) as a solid line and an $\ell=2$
quasinormal mode fit from run {\it r1}.  The long and short dashed line
is the offset from sphericity, determined by the fit, showing that the
black hole is oscillating about a Kerr hole with rotation parameter
$a/m=.35$.}
\label{fig:cr,r1}
\end{figure}

\begin{figure}
\caption{This figure shows the numerically extracted $\ell=2$
waveform and its fit to the two lowest $\ell=2$ quasinormal modes for
the run labeled {\it r2}.}
\label{fig:l=2,r2}
\end{figure}

\begin{figure}
\caption{This figure shows the numerically extracted $\ell=3$
waveform and its fit to the two lowest $\ell=3$ quasinormal modes for
the run labeled {\it r2}.}
\label{fig:l=3,r2}
\end{figure}

\begin{figure}
\caption{This figure shows the numerically extracted $\ell=4$
waveform and its fit to the two lowest $\ell=4$ quasinormal modes for
the run labeled {\it r2}.}
\label{fig:l=4,r2}
\end{figure}

\begin{figure}
\caption{This figure shows the numerically extracted $\ell=5$
waveform and its fit to the two lowest $\ell=5$ quasinormal modes for
the run labeled {\it r2}.}
\label{fig:l=5,r2}
\end{figure}

\begin{figure}
\caption{This figure shows the numerically extracted $\ell=2$
waveform and its fit to the two lowest $\ell=2$ quasinormal modes for
the run labeled {\it r4}.}
\label{fig:l=2,r4}
\end{figure}

\begin{figure}
\caption{This figure shows the numerically extracted $\ell=3$
waveform and its fit to the two lowest $\ell=3$ quasinormal modes for
the run labeled {\it r4}.}
\label{fig:l=3,r4}
\end{figure}

\begin{figure}
\caption{This figure shows the numerically extracted $\ell=4$
waveform and its fit to the two lowest $\ell=4$ quasinormal modes for
the run labeled {\it r4}.}
\label{fig:l=4,r4}
\end{figure}

\begin{figure}
\caption{This figure shows the numerically extracted $\ell=5$
waveform and its fit to the two lowest $\ell=5$ quasinormal modes for
the run labeled {\it r4}.}
\label{fig:l=5,r4}
\end{figure}

\begin{table}
\begin{tabular}{||d||d|d||} \hline
Case & $J$ & $a/m$ \\ \hline \hline
k1 & 2.5 & .481 \\ \hline
k2 & 5.0 & .677 \\ \hline
\end{tabular}
\caption{
This table gives summary data for evolutions of pure Kerr spacetimes.
$J$ is the total angular momentum of the spacetime, $a/m$ is the
usual rotation parameter in the Kerr metric.  }
\label{tab:the_k's}
\end{table}

\begin{table}
\begin{tabular}{||d||d|d|d|d|d|d||} \hline
Case & $J$ & $Q_0$ & $\sigma$ & $\eta_0$ & n &
$n^\prime$ \\ \hline \hline
r0  & 0.0 & 1.0 & 1.0 & 1.0 & 2  & N/A \\ \hline
r1  & 2.5 & 1.0 & 1.0 & 1.0 & 2 & N/A \\ \hline
r2  & 5.0 & 1.0 & 1.0 & 1.0 & 2  & N/A  \\ \hline
r3  & 10.0 & 1.0 & 1.0 & 1.0 & 2 & N/A  \\ \hline
r4  & 10.0 & 0.5 & 1.0 & 0.0 & 2 & N/A  \\ \hline
r5  & 15.0 & 0.0 & N/A & N/A & N/A & N/A  \\ \hline
\hline
o1  & 0.0 & 2.0 & 1.0 & 2.0 & N/A & 3 \\ \hline
o2  & 0.0 & 2.0 & 1.0 & 2.0 & N/A & 5  \\ \hline
\end{tabular}
\caption{This table gives input parameters for each of the runs.
$J$ describes the amount of angular momentum in the system, $Q_0$ the
amplitude of the Brill wave placed in the spacetime, $\sigma$ gives
the width of the Brill wave, gives $\eta_0$ the location of the peak of
the Brill wave distortion, and the parameters $n$ ($n^\prime$)
describe the radial dependence of the distortion in the metric
(extrinsic curvature).}
\label{tab:initial}
\end{table}

\begin{table}
\begin{tabular} {||d||d|d|d||} \hline
Case &
$(a/m)_{min}$ &
$(a/m)_{max}$ &
$(a/m)_{cr}$  \\ \hline
r0 & .000 & .000 & .000 \\ \hline
r1 & .342 & .600 & .351 \\ \hline
r2 & .432 & .698 & .436 \\ \hline
r3 & .511 & .759 & .512 \\ \hline
r4 & .696 & .827 & .703 \\ \hline
r5 & .868 & .886 & .870 \\ \hline
\hline
k1 & .480 & .481 & .481 \\ \hline
k2 & .675 & .677 & .677 \\ \hline
\end{tabular}
\caption{This is a table of $a/m$ values calculated from the initial data.
$(a/m)_{min} = J/M^2_{ADM}$ on the initial slice and will be the final
value if all the energy in the spacetime goes into the horizon, and
$(a/m)_{max}=J/M_{AH}^2$ on the initial slice which will be the final
value if the surface area of the horizon does not increase during the
evolution.  The parameter $(a/m)_{cr}$ is the angular momentum
extracted from the horizon.}
\label{tab:a/m}
\end{table}

\begin{table}
\begin{tabular} {||d||d|d|d|d|d||}
Case &
$ M_{rad}^{\ell=2} $ &
$ M_{rad}^{\ell=3} $ &
$ M_{rad}^{\ell=4} $ &
$ M_{rad}^{\ell=5} $ &
$ M_{rad}^{total}/M_{ADM} $ \\ \hline
\hline
r0& 94.0\% & 0.0\% & 6.0\% & 0.0\% & 8.37e-2 \\ \hline
r1& 94.9\% & 2.0\% & 2.8\% & .3\% & 3.40e-2 \\ \hline
r2& 94.8\% & 3.4\% & 1.5\% & .3\% & 1.84e-2 \\ \hline
r3& 93.9\% & 5.2\% & .7\% & .2\% & 9.41e-3 \\ \hline
r4& 92.1\% & 7.0\% & .6\% & .2\% & 8.30e-3 \\ \hline
r5& 95.7\% & 4.1\% & .2\% & .02\% & 1.00e-3 \\ \hline
\hline
o1& .05\% & 99.9\% & .02\% & 0.0001\% & 1.06e-4 \\ \hline
o2& .06\% & 78.9\% & 0.00\% & 21.0\% & 2.88e-4 \\ \hline
\hline
k1& N/A & N/A & N/A & N/A & 4.62e-8 \\ \hline
k2& N/A & N/A & N/A & N/A & 2.00e-7 \\ \hline
\end{tabular}
\caption{This table shows the total energy emitted as radiation
divided by the ADM mass in the right-most column, and it shows
the fraction of this energy emitted in the first four quasinormal
modes of the black hole spacetime.}
\label{tab:more_masses}
\end{table}

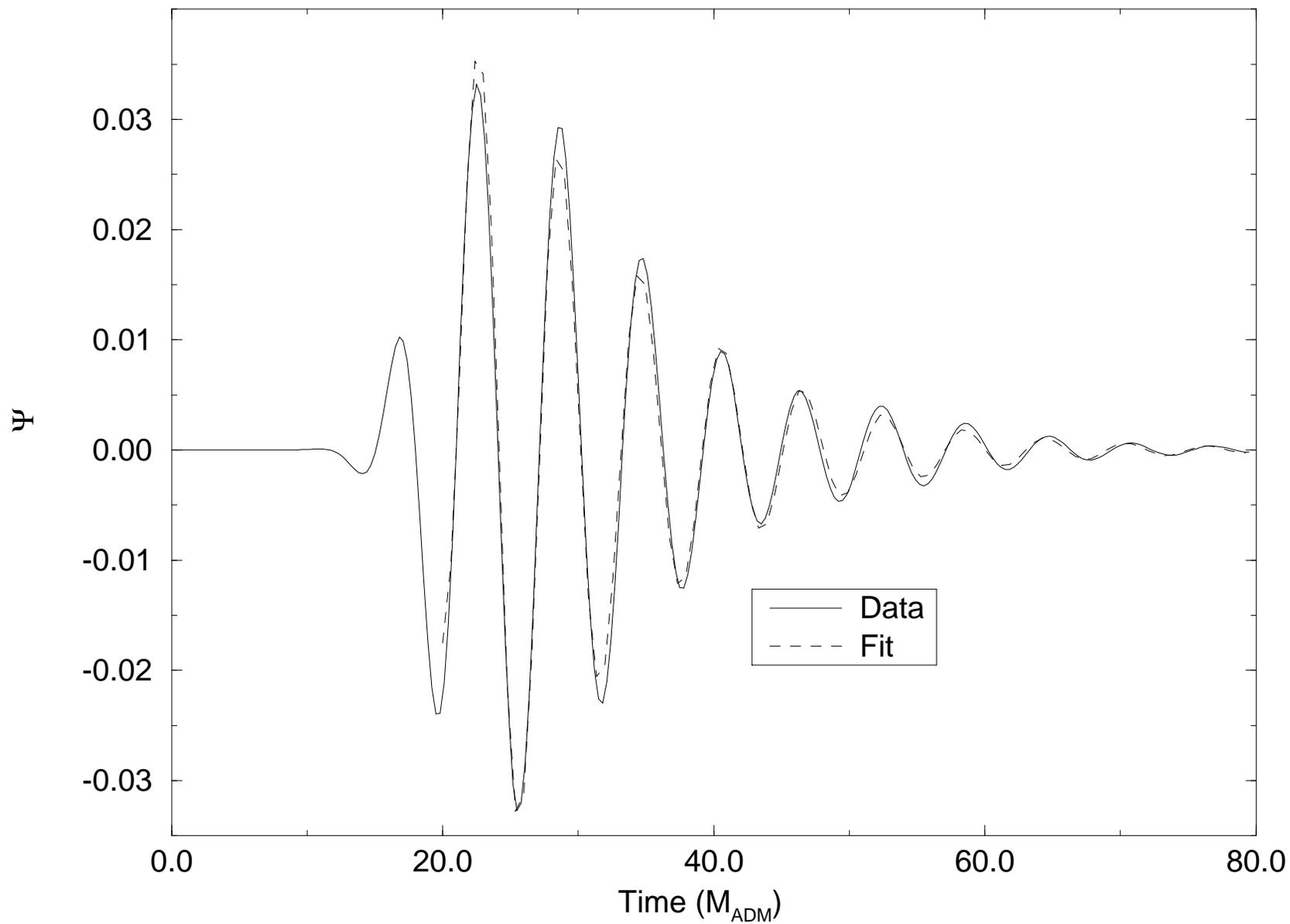
\end{document}